\documentclass[aps,prb,reprint,longbibliography,superscriptaddress]{revtex4-2}
\usepackage{mathrsfs}
\usepackage{amsmath,gensymb}
\usepackage{amsfonts}
\usepackage{amssymb}
\usepackage{amsthm}
\usepackage{graphicx}
\usepackage{natbib}
\usepackage{color}
\usepackage{hyperref}
\usepackage{bm}
\usepackage[caption=false]{subfig}
\usepackage{verbatim}
\usepackage[normalem]{ulem}

\begin{document}
\title{Ultra-strong coupling of two ferromagnets via Meissner currents}

\author{V. M. Gordeeva}
\affiliation{Moscow Institute of Physics and Technology, Dolgoprudny, 141700 Moscow region, Russia}

\author{G. A. Bobkov}
\affiliation{Moscow Institute of Physics and Technology, Dolgoprudny, 141700 Moscow region, Russia}

\author{A. M. Bobkov}
\affiliation{Moscow Institute of Physics and Technology, Dolgoprudny, 141700 Moscow region, Russia}

\author{I. V. Bobkova}
\affiliation{Moscow Institute of Physics and Technology, Dolgoprudny, 141700 Moscow region, Russia}
\affiliation{National Research University Higher School of Economics, 101000 Moscow, Russia}

\author{Tao Yu}
\affiliation{School of Physics, Huazhong University of Science and Technology, Wuhan 430074, China}

\begin{abstract}
In this work, we study the magnetization dynamics in a ferromagnet/insulator/ferromagnet trilayer sandwiched between two superconductors (S/F/I/F/S heterostructure). It is well-known that a conceptually similar S/F/S system is a platform for implementing ultra-strong magnon-photon coupling. Here, we demonstrate that in such S/F/I/F/S heterostructure, ultra-strong magnon-magnon
coupling between the two F layers also appears. The strength of this interaction is many times greater than the strength of the usual dipole-dipole interaction. It is mediated via Meissner currents excited in the superconductor layers by the magnon stray fields. The strength of the magnon-magnon
coupling is anisotropic, and its anisotropy is opposite to the anisotropy of the magnon-photon coupling, which allows them to be separated. Both couplings become much stronger when the temperature drops below the critical temperature of the superconductor layers. It enables the implementation of an efficient tuning of the wavenumber in the S/F/I/F/S heterostructures controlled by temperature in a wide range of frequencies. Overall, the rich and tunable spectrum of S/F/I/F/S multilayers opens broad prospects for their application in magnonics.

\end{abstract}

\maketitle

\section{Introduction}
\label{intro}
In recent years different concepts of magnonic logic and signal processing have been proposed~\cite{Chumak2015,Chumak2014,Lee2008,Schneider2008,Klingler2014,Ganzhorn2016,Dutta2015,Klingler2015,Nikitin2015,Sato2013}. One of the most important issues of the magnonic technology is an efficient, controllable and reconfigurable  connection of separate magnonic signal processing devices into a magnonic circuit.  In particular, 
magnetic stripes coupled by the dipolar coupling~\cite{Sadovnikov2015,Wang2018} and  artificial materials---magnonic crystals \cite{Nikitov2001,Chumak2009,Gubbiotti2010,Nikitin2017,Nikitin2020,Nikitin2023}---were proposed to realize a controlled connection between magnonic conduits. For engineering of magnonic networks the ability to on-chip modulate and tune the spin wave dispersion and, in particular, the coupling strength of the magnetic stripes, is one of the most important requirements.

Different ways to control and tune the spin wave dispersion were proposed. In particular, the interlayer exchange coupling in ferromagnet/normal metal/ferromagnet (F/N/F) heterostructures was found to strongly modify the dispersion relation of the ferromagnets~\cite{Wigen1993}. The tunability of the spin wave
characteristics can be achieved by changing the bias magnetic field; in layered structures that contain both ferromagnetic and ferroelectric layers it is possible to maintain
a dual electric and magnetic control \cite{Nikitin2017_2,Nikitin2017_3}. In this respect, ferromagnetic heterostructures composed of ferromagnets
(Fs) and normal metals (Ns) were also intensively explored \cite{Demidov2002,Qiu2024}.  The dipolar fields emitted by the spin waves drive the diamagnetic
currents in the Ns. Thus, the adjacent metal works as a spin sink strongly influencing the Gilbert damping of the magnon modes~\cite{Tserkovnyak2005,Bunyaev2020,Bertelli2021,Nikitin2018,Gladii2019,Kostylev2016,Ye2024}.

When the normal metals become superconducting, additional functionalities appear. Two mechanisms of interaction between spin waves and superconductors (Ss) were discussed in the literature: exchange and electromagnetic ones. It was reported that the interface exchange coupling between the electrons in the superconductor and magnetization in the ferromagnet or an antiferromagnet (AF) results in the emergence of composite particles composed of a magnon in F (AF) and an accompanying cloud of spinful triplet Cooper pairs in S~\cite{Bobkova2022,Bobkov2023}. The other consequence of such exchange coupling is the coupling of magnons of two Fs in F/S/F heterostructures mediated by the spin supercurrents \cite{Ojajarvi2022}.

The electromagnetic interaction between the Fs and Ss results in the appearance of skyrmion-fluxon and magnon-fluxon excitations \cite{Hals2016,Baumard2019,Dahir2019,Andriyakhina2021,Bihlmayer2021,Menezes2019,Petrovic2021,Dobrovolskiy2019}. In 
superconductors spin waves  also generate stray magnetic
fields that shift the spin wave frequency \cite{Li2018,Jeon2019,Golovchanskiy2020,Mironov2021,Silaev2022,Kuznetsov2022,Zhou2023}. Thus, local superconductor gates on the magnetic films can play a role of potential barriers reflecting propagating spin waves \cite{Volkov2009,Golovchanskiy2018,Golovchanskiy2018_2,Golovchanskiy2019,Golovchanskiy2020_2,Borst2023}.  The Meissner screening can also increase the magnon group velocity thus enhancing the magnon transport \cite{Zhou2024} and mediate the coupling between ferromagnetic layers resulting in the formation of antiferromagnetic interaction between the F-layers \cite{Golovchanskiy2023}. The important property of S/F heterostructures is that at low temperatures the Ohmic
dissipation produced by the induced eddy currents disappears. The other important property of the S layers is that they lead to an ultrastrong coupling \cite{Qin2024,Kockum2019} between magnons and polaritons in S/F/S heterostructures in the microwave cavities, forming magnon-polariton modes, which was reported both theoretically and experimentally \cite{Golovchanskiy2021,Golovchanskiy2021sa,Silaev2023}.  The mechanism of such ultrastrong coupling is also based on the modulation of the electromagnetic field by the Meissner currents induced in the superconductors and the strong
confinement of microwave magnetic field across the heterostructure. Furthermore, a strong
anisotropy of this coupling was reported \cite{Qiu2024}. Its coupling strength is comparable to the bare magnon frequency for 
the bulk volume configuration, that is, when the excitation wave vector $\bm k$ is aligned with the equilibrium magnetization $\bm M$ of the F layer, but vanishes in the Damon-Eshbach case, when $\bm k \perp \bm M$.

In this work, we consider a ferromagnet/insulator/ferromagnet trilayer sandwiched between two Ss (S/F/I/F/S). We demonstrate that this system, which is conceptually very similar to the one investigated in Refs.~\onlinecite{Silaev2023,Qiu2024},  is the simplest system, where not only ultrastrong magnon-photon, but also ultrastrong magnon-magnon coupling between magnons in spatially separated ferromagnetic layers appears. Its mechanism is also based on the giant enhancement of the demagnetization fields in the F layers produced by the Meissner currents flowing
in the S layers.  In the range of magnon wave numbers of the order of microwave photons $k \lesssim 1000 \rm{m^{-1}}$, the strength of this interaction is many times larger than the strength of the usual dipole-dipole interaction between ferromagnetic layers having thicknesses under consideration (tens of nanometers) and located at a distance of several tens of nanometers. The strength of the magnon-magnon coupling is also anisotropic. It is interesting that the anisotropy of the magnon-magnon coupling is opposite to the anisotropy of the magnon-photon coupling, i.e., its strength is maximal in the Damon-Eshbach configuration and tends to zero at large $k$ when $\bm k$ is aligned with $\bm M$. The anisotropy thereby allows to separate magnon-magnon and magnon-photon couplings. We also demonstrate the possibility of controlling magnon-magnon as well as magnon-photon couplings by adjusting the temperature. Both couplings become much stronger when the temperature drops below the superconducting transition temperature of the S layers. It allows to implement an effective tuning of the wave number in the S/F/I/F/S heterostructures controlled by temperature in a wide range
of frequencies of the order of tens of GHz, which should be important for applications requiring essential and easily controllable phase shifts of propagating excitations.

\section{system}
\label{system}

Here, we consider a superconductor/ferromagnet/insulator/ferromagnet/superconductor (S/F/I/F/S) heterostructure hosting both the photonic modes and magnons. The ferromagnets F can be both insulating (FI) or metallic (FM). The insulating layer I between the ferromagnets is needed only to prevent direct interface exchange interaction between them. The F/I/F trilayer is sandwiched between two thick superconductors S with thicknesses $\gg \lambda$, where $\lambda$ is London's penetration depth of the superconductors for the magnetic field. Following the existing experimental implementations of the similar S/F/I/S systems \cite{Golovchanskiy2021,Golovchanskiy2021sa}, we assume $\lambda \sim 80-100$~nm and $\{d_F, d_I\} \lesssim \lambda$.  Since the typical wave numbers of microwave photons $k \sim 0.1 $ to $1~{\rm mm^{-1}}$, in the case under consideration the approximation $k \lambda \ll 1$ works very well.

\begin{figure}[tb]
\begin{center}
\includegraphics[width=90mm]{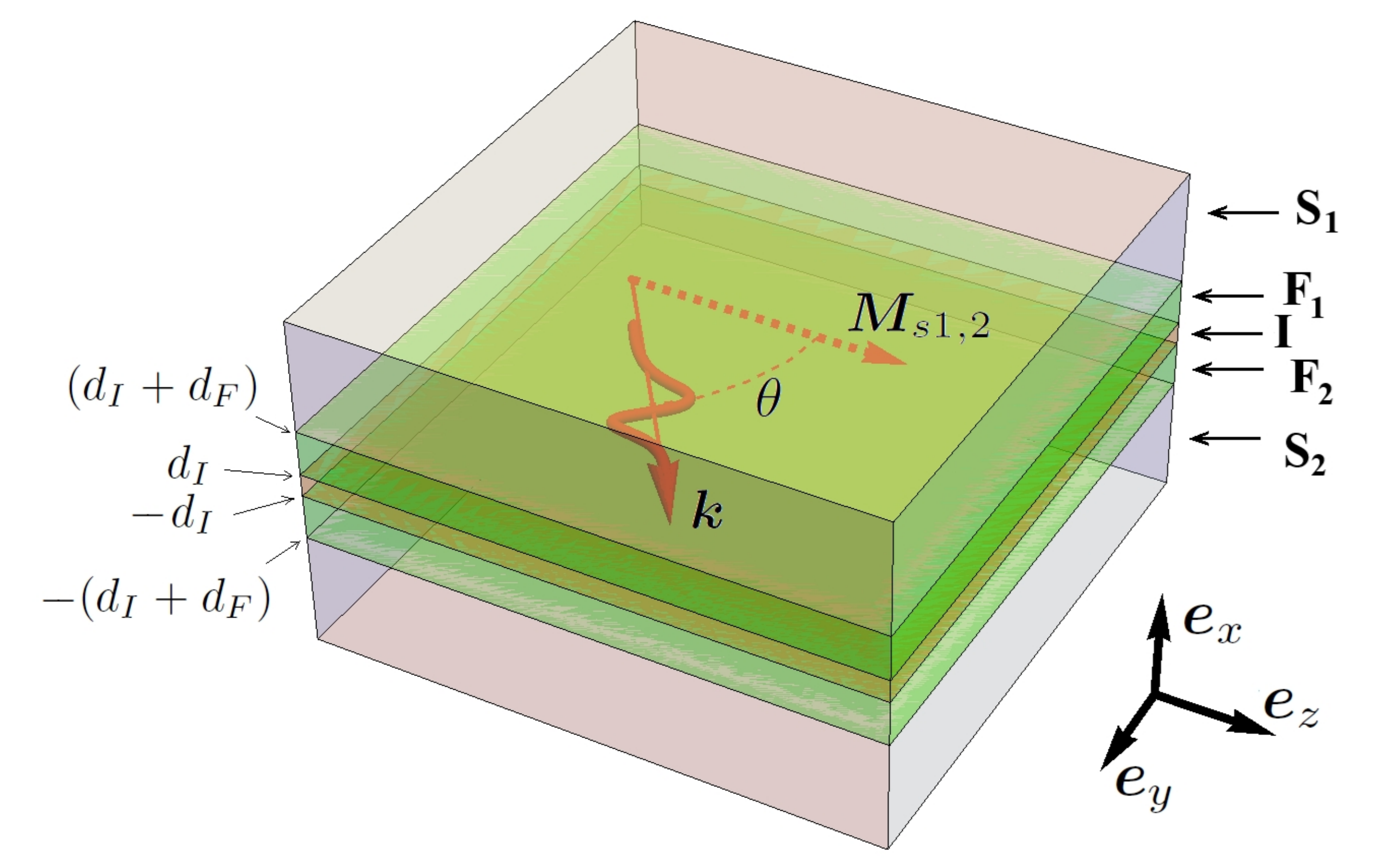}
\caption{Sketch of the F/I/F trilayer sandwiched between two S layers. The equilibrium magnetizations of the $\rm F_{1,2}$ layers $\bm M_{s1,2}$ are aligned with the applied magnetic field $H_0 \bm e_z$. The magnon with wavevector $\bm k$ can propagate at an angle $\theta$ with respect to $\bm M_{s1,2}$.}
\label{fig:sketch}
\end{center}
\end{figure}

The sketch of the system is presented in Fig.~\ref{fig:sketch}. The coordinate system is defined such that the $x$ axis is perpendicular to the layers' interfaces with $x=0$ in the middle of the I layer. The insulator has a thickness $2d_I$, and for simplicity, we assume that both ferromagnets have the same thicknesses $d_{F}$ since considering the case with different thicknesses does not influence the physical picture of the studied effects, at least until the condition $\{d_F, d_I\} \lesssim \lambda$ is violated.   
The in-plane static external magnetic field $\bm{H}_0=H_0 \bm e_z$ biases the equilibrium magnetizations $\bm M_{s1}=M_1\bm e_z $ and $\bm M_{s2}=M_2\bm e_z $ in the F1 and F2 layers, respectively. We treat magnons in the linear response regime, when they can be described by the transverse fluctuations of the magnetizations in F$_1$ and F$_2$  as $\bm M_{\omega 1(2)}(\bm \rho, t)=\tilde {\bm M}_{1(2)} e^{i{\bm k}\cdot{\bm \rho}-i\omega t}$ with amplitudes $\tilde {\bm M}_{1(2)}=\tilde M_{1(2)x}\bm e_x+\tilde M_{1(2)y}\bm e_y$, wavevector $\bm k=k_y \bm e_y+k_z \bm e_z$, in-plane radius vector $\bm \rho=y \bm e_y+z \bm e_z$, and frequency $\omega$. The full magnetizations in F1 and F2 layers are $\bm M_{full,1(2)}=\bm M_{s1(2)}+\bm M_{\omega 1(2)}$. $\theta$ is the angle between the wave vector $\bm k$ and the equilibrium magnetizations $\bm M_{s1}$ and $\bm M_{s2}$, i.e., $k_z=k\cos\theta$ and $k_y=k\sin\theta$ with $k=|\bm k|$.

\section{Ultra-strong magnon-magnon and magnon-photon coupling}
\label{classical}

\subsection{Method}
\label{method}

Magnetization dynamics radiates electric and magnetic fields, which induce the Meissner currents in the S layers and are strongly renormalized by these currents. In its turn, such renormalization of the stray fields strongly influences the magnon and photon dispersion relations. We derive the dispersion relations of the magnonic and photonic excitations by solving the coupled system of Maxwell equations for electric and magnetic fields and the Landau-Lifshitz-Gilbert (LLG) equation for the magnetizations. Maxwell equations for the electric field $\bm E(\bm r,t)$, magnetic field $\bm H(\bm r,t)$, magnetic field induction $\bm B(\bm r,t)$, and electric current $\bm J(\bm r,t)$ take the form
\begin{align}
    &\mathrm{rot}~ \bm E=-\dfrac{\partial \bm B}{\partial t},\nonumber\\
    &\mathrm{rot}~ \bm H=\bm J+\varepsilon \dfrac{\partial \bm E}{\partial t},
    \label{Maxwell}
\end{align}
where $\varepsilon$ is the dielectric constant of the corresponding layers. Considering $\bm M_{\omega}(\bm \rho, t)=\tilde {\bm M} e^{i\bm k \cdot \bm \rho-i\omega t}$ (for now we omit the indices 1 or 2 in the magnetization expression in order to obtain general equations valid for all layers), we assume $\bm E(\bm r,t)=\tilde {\bm E}(x)e^{i\bm k \cdot \bm \rho-i\omega t}$. As $\bm B=\mu_0(\bm H+\bm M)$, where $\bm M \equiv \bm M_{full}$, and $\bm J=\sigma \bm E$, where $\mu_0$ is the vacuum magnetic permeability and $\sigma$ is the conductivity of the layer, from Eq.~(\ref{Maxwell}) we obtain the equation of motion for the electric field $\bm E$, which takes the following form in the different layers of the system:
\begin{align}
    &\mathrm{in ~S:}~~\Delta \bm E+k_S^2 \bm E=0,~~k_S=\sqrt{\omega^2 \mu_0\varepsilon_0-\dfrac{1}{\lambda^2}},\nonumber\\
     &\mathrm{in ~F:}~~\Delta \bm E+k_F^2 \bm E=-i\omega \mu_0 \mathrm{rot}\bm M,\nonumber\\
     & k_F=\begin{cases}
\omega\sqrt{\mu_0\varepsilon_{FI}}, ~\mathrm{in ~FI}\\
\sqrt{\omega^2 \mu_0\varepsilon_0+i\omega\mu_0\sigma_{FM}}, ~\mathrm{in ~FM}
     \end{cases},\nonumber\\
    &\mathrm{in ~I:}~~\Delta \bm E+k_I^2 \bm E=0, ~~k_I=\omega\sqrt{\mu_0\varepsilon_{I}}.
    \label{Maxwell_final}
\end{align}
Here, $\varepsilon_0$, $\varepsilon_{FI}$, and $\varepsilon_I$ are the dielectric constants of the vacuum, FI, and I, respectively, $\sigma_{FM}$ is the conductivity of the metallic ferromagnets (we assume the same constants describing the electric qualities for both ferromagnets). In the framework of the two-fluid model, the conductivity of a superconductor at frequency $\omega$ takes the form \cite{Schmidt_book}:
\begin{align}
\sigma_S (\omega) = \frac{\rho_n e^2 \tau}{m_e(1+\omega^2 \tau^2)} + i \Bigl(  \frac{\rho_n e^2 \omega \tau^2}{m_e(1+\omega^2 \tau^2)} + \frac{\rho_s e^2 }{m_e \omega}\Bigr),   
    \label{conductivity_S}
\end{align}
where $\rho_s = \rho(1-(T/T_c)^4)$ and $\rho_n = \rho (T/T_c)^4$ are superfluid and normal fluid densities, respectively. $\rho$ is the electron density, $m_e$ is the electron mass, $e$ is the electron charge, $T_c$ is the critical temperature of the superconductor, $T$ is the temperature of the system, and $\tau$ is the relaxation time of electrons. In typical metals $\tau \sim 10^{-12}$~s and, therefore, at the microwave frequencies $\omega \lesssim 100$~GHz we have $\omega \tau \ll 1$. Then $\sigma_S(\omega)$ can be simplified as 
\begin{align}
\sigma_S (\omega) = \sigma_n \left(\frac{T}{T_c}\right)^4 + \frac{i}{\omega \mu_0 \lambda^2},   
    \label{conductivity_S_approximate}
\end{align}
where $\sigma_n = \rho e^2 \tau/m_e$ is the conductivity of the normal metals and $\lambda = \sqrt{m_e/(\mu_0 \rho_s e^2)}$ is the London’s
penetration depth. Taking into account that the conductivity $\sigma_n \sim 10^6$ to $10^7 ~\rm{\Omega^{-1}m^{-1}}$ we obtain that the first term in Eq.~(\ref{conductivity_S_approximate}) can be safely disregarded with respect to the second one at moderate temperatures, not very close to $T_c$, and the conductivity of the superconductors can be taken as $\sigma_S=i/(\omega\mu_0\lambda^2)$.  

Solving Eq.~(\ref{Maxwell_final}) in each layer, we obtain the following expressions for the amplitudes $\tilde {\bm E}(x)$:
\begin{align}
    &\mathrm{in ~S1:}~~\tilde {\bm E}(x)=\bm S_1 e^{iB(x-d_I)},\nonumber\\
     &\mathrm{in ~S2:}~~\tilde {\bm E}(x)=\bm S_2 e^{-iB(x+d_I)},\nonumber\\
     &\mathrm{in ~F1:}~~\tilde {\bm E}(x)=\bm F_1^+e^{iC(x-d_I)}+\bm F_1^- e^{-iC(x-d_I)}\nonumber\\
     &+\dfrac{\omega\mu_0}{C^2}\bm k\times\tilde {\bm M}_1,\nonumber\\
     &\mathrm{in ~F2:}~~\tilde {\bm E}(x)=\bm F_2^+e^{iC(x+d_I)}+\bm F_2^- e^{-iC(x+d_I)}\nonumber\\
     &+\dfrac{\omega\mu_0}{C^2}\bm k\times\tilde {\bm M}_2,\nonumber\\
     &\mathrm{in ~I:}~~\tilde {\bm E}(x)=\bm I^+e^{iAx}+\bm I^- e^{-iAx},
\label{Maxwell_solutions}
\end{align}
where $A=\sqrt{k_I^2-k^2}$, $B=\sqrt{k_S^2-k^2}$, $C=\sqrt{k_F^2-k^2}$, and $\{\bm S_{1,2}, \bm F_{1,2}^{\pm}, \bm I^{\pm} \}$ are vectors of unknown coefficients, which can be explicitly written as $\bm S_1=(S_{1x}, S_{1y}, S_{1z})^T$ and similarly for other vectors. Upon deriving Eq.~(\ref{Maxwell_solutions}), it is assumed that the F layers are thin enough such that their magnetizations $\bm M_{full,1(2)}$ can be considered constant throughout the entire thickness of the corresponding layer along the normal $x$-axis.

From the first equation in (\ref{Maxwell}) we can express the magnetic field components via $\bm E$ and $\bm M$:
\begin{align}
    &H_x=\dfrac{k_yE_z-k_zE_y}{\omega\mu_0}-M_x,\nonumber\\
     &H_y=\dfrac{ik_zE_x-\partial_xE_z}{i\omega\mu_0}-M_y,\nonumber\\
     &H_z=\dfrac{\partial_xE_y-ik_yE_x}{i\omega\mu_0}.
     \label{H_comp}
\end{align}
The $x$-component of the second equation in (\ref{Maxwell}) gives us 
\begin{align}
    E_x=\dfrac{i(k_yH_z-k_zH_y)}{\sigma-i\varepsilon\omega}.
    \label{E_x}
\end{align}
In the S layers we have $|k_ik_j/i\omega\mu_0 (\sigma-i\varepsilon\omega)|=|k_ik_j/k_S^2| \sim (k\lambda)^2 \ll 1$ for $i,j\in\{y,z\}$. Substituting Eq.~(\ref{E_x}) into Eq.~(\ref{H_comp}) and taking into account the smallness of the above parameter, we obtain for $H_y$ and $H_z$ in different layers:
\begin{align}
    &\mathrm{in ~S:}~~H_y=-\dfrac{\partial_xE_z}{i\omega\mu_0}, ~~H_z=\dfrac{\partial_xE_y}{i\omega\mu_0},\nonumber\\
    &\mathrm{in ~I:} ~~\left(
\begin{array}{cc}
H_y\\H_z
\end{array}
\right)=\dfrac{1}{A^2}\hat K_I\left(\begin{array}{cc}
-\dfrac{\partial_xE_z}{i\omega\mu_0}\\\dfrac{\partial_xE_y}{i\omega\mu_0}
\end{array}
\right),\nonumber\\
&\mathrm{in ~F:} ~~\left(
\begin{array}{cc}
H_y\\H_z
\end{array}
\right)=\dfrac{1}{C^2}\hat K_F\left(\begin{array}{cc}
-\left(\dfrac{\partial_xE_z}{i\omega\mu_0}+M_y\right)\\\dfrac{\partial_xE_y}{i\omega\mu_0}
\end{array}
\right),
\label{H_final}
\end{align}
where
\begin{align}
\hat K_{I(F)}=\left(
\begin{array}{cc}
\beta_{I(F)} & -\dfrac{K_1}{2} \\  -\dfrac{K_1}{2}
 & \alpha_{I(F)}
\end{array}
\right),
\end{align}
$\alpha_{I(F)}=k_{I(F)}^2-k^2\cos^2\theta$,  $\beta_{I(F)}=k_{I(F)}^2-k^2\sin^2\theta$, and  $K_1=k^2\sin2\theta$. Here, $\theta$ is the direction ${\bf k}$ of the mode propagation with respect to the saturation magnetization direction ${\bf M}_0$.

The Maxwell equations (\ref{Maxwell}) are accompanied by the boundary conditions, which are the continuity of $E_y, E_z, H_y$, and $H_z$ at all the four interfaces $x=\pm d_I$ and $x=\pm(d_I+d_{F})$. Since we assume that $\{d_I, d_{F}\}\lesssim 100 $~nm, $k< 10^4$~m$^{-1}$, $\omega\lesssim10$~GHz, $\varepsilon_{I, FI}\sim10 \varepsilon_0$, and $\sigma_{FM}\sim10^7 
~\Omega^{-1}\cdot $m$^{-1}$, therefore $\{|Ad_I|, |Cd_{F}|\}\ll 1$. Matching $E_y, E_z, H_y$ and $H_z$ described by Eqs.~(\ref{Maxwell_solutions}) and (\ref{H_final}) at all the interfaces, one can express all the fields via the magnon amplitudes $\tilde {\bm M}_{1,2}$. In the leading order with respect to the small parameters $|Ad_I|$ and $|Cd_{F}|$, we obtain the following expressions for the demagnetization field produced by the magnetic excitations in the ferromagnets (the details of the derivation are presented in the Appendix): 
 \begin{align}
     \hat H=-\hat N \hat {\tilde M}, ~~\hat H=\left( \begin{array}{c}
          \tilde H_{x1}  \\
          \tilde H_{x2}\\
          \tilde H_{y1}\\
          \tilde H_{y2}
     \end{array}\right),~~\hat {\tilde M}=\left( \begin{array}{c}
          \tilde M_{1x}  \\
          \tilde M_{2x}\\
          \tilde M_{1y}\\
          \tilde M_{2y}
     \end{array}\right),
     \label{H_M}
 \end{align}
where we have introduced the amplitudes $\tilde H_{x,y1(2)}$ so that $\bm H(\bm r,t)=\tilde {\bm H}(x)e^{i\bm k \cdot \bm \rho-i\omega t}$, the indices ``1" (``2") denote the magnetic field in F1 and F2, respectively. Up to the leading order with respect to  $\{|Ad_I|, |Cd_{F}|\}$, the $x$-dependence of the demagnetization fields $\tilde {\bm H}_{1,2}$ can be disregarded. $\hat N$ is the demagnetization tensor, which takes the form:
 \begin{align}
     &\hat N=\left(\begin{array}{cccc}
        1  & 0&0&0 \\
         0 &1&0&0\\
         0&0&N&N\\
         0&0&N&N
     \end{array}\right),\nonumber\\
     &N=\dfrac{Bd_{F}(k_I^2 k_F^2\alpha-Bk^2\chi\sin^2\theta)}{2\alpha(Bk^2\chi-k_I^2 k_F^2\alpha)},
     \label{N_final}
 \end{align}
where $\alpha=i+B(d_F+d_I)$, $\chi=d_Fk_I^2+d_Ik_F^2$, and for the considered parameters $B \approx i/\lambda$ with very good accuracy. 

Now, we turn to the LLG equation
 \begin{align}
     \dfrac{\partial\bm M}{\partial t}=-\gamma\mu_0\bm M\times\bm H + \frac{\alpha}{M_{1(2)}} \bm M \times \frac{\partial\bm M}{\partial t},
     \label{LLG}
 \end{align}
 where $-\gamma$ is the gyromagnetic ratio of electrons and $\alpha$ is the Gilbert damping constant. By linearizing this equation with respect to the magnon amplitudes and the magnon demagnetization fields $\tilde {\bm H}_{1,2}$, we obtain
\begin{align}
     &i\omega\tilde M_{1(2)x}=\gamma\mu_0(\tilde M_{1(2)y}H_0-M_{1(2)}\tilde H_{y1(2)})- i \alpha \omega \tilde M_{1(2)y},\nonumber\\
     &i\omega\tilde M_{1(2)y}=\gamma\mu_0(M_{1(2)}\tilde H_{x1(2)}-\tilde M_{1(2)x} H_0) + i \alpha \omega \tilde M_{1(2)x}.
     \label{LLG_comp}
 \end{align}
 After substituting $\hat H$ from Eq.~(\ref{H_M}), Eq.~(\ref{LLG_comp}) can be rewritten as
 \begin{widetext}
 
  \begin{align}
     \left(\begin{array}{cccc}
          i\omega&0&-\gamma\mu_0 (H_0+M_1N) + \alpha i\omega&-\gamma\mu_0 M_1N \\
         0 & i\omega&-\gamma\mu_0 M_2N&-\gamma\mu_0 (H_0+M_2N)+ \alpha i\omega\\
         \gamma\mu_0 (H_0+M_1)-\alpha i\omega&0&i\omega&0\\
         0&\gamma\mu_0 (H_0+M_2)-\alpha i\omega&0&i\omega
     \end{array}\right)\hat{\tilde M}=0.
     \label{M_matrix}
 \end{align}
    \end{widetext}
The eigenfrequencies of the S/F/I/F/S heterostructure are to be obtained from Eq.~(\ref{M_matrix}).

\subsection{Anisotropic giant magnon splitting and magnon-polaritons. Identical F-layers.}
\label{symmetric_classical}

\begin{figure*}
    \centering
    \includegraphics[width=\linewidth]{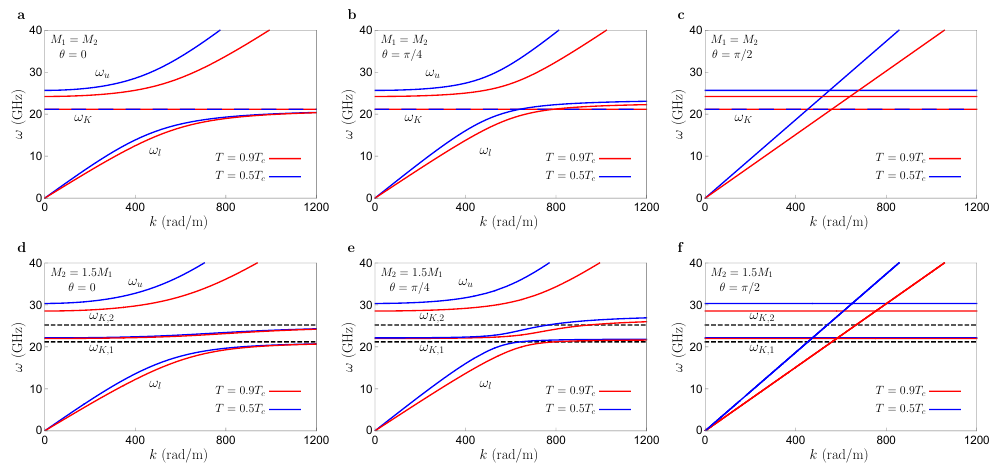}
    \caption{(a)-(c) Frequency dispersion of the eigenmodes in the S/F/I/F/S structure in the symmetric configuration with $M_1=M_2$ at different mode propagation directions $\theta$. Different curves in each panel correspond to different temperatures. The important difference from the results of Ref.~\onlinecite{Qiu2024} is the presence of the additional optical mode $\omega_K$ (shown by red and blue line), which does not depend on temperature and does not manifest anisotropy, see text. (d)-(f) The same as in (a)-(c), but in the asymmetric case with $M_2=1.5M_1$. The dashed black lines correspond to $T>T_c$, that is, the limit $\lambda\to\infty$, and represent the Kittel modes of the non-interacting ferromagnets. $d_F=10$ nm, $d_I = 10$~nm, $\lambda(T=0) = 80$ nm, $\mu_0M_0=\mu_0M_1=0.24$~T, $\mu_0H_0=50$~mT,  $\varepsilon_{FI}=\varepsilon_I = 8\varepsilon_0$, and $\gamma=1.76\cdot10^{11}$~Hz$\cdot\mathrm{T}^{-1}$.}
    \label{fig:dispersion}
\end{figure*}

At first we discuss the symmetric structure with $M_1 = M_2=M_0$. In this case, Eq.~(\ref{M_matrix}) results in the following eigenfrequencies
\begin{align}
&\omega_1=\gamma\mu_0\sqrt{H_0(H_0+M_0)},\label{modes_symm_1} \\
&\omega_2=\gamma\mu_0\sqrt{(H_0+2M_0N)(H_0+M_0)},
\label{modes_symm_2}
\end{align}
where $\omega_1$ corresponds to the Kittel mode frequency $\omega_K$ of an isolated ferromagnet \cite{Kittel}. In Eqs.~(\ref{modes_symm_1})-(\ref{modes_symm_2}) the corrections due to the Gilbert damping $\sim \alpha^2 $ are disregarded. The first-order corrections account for the decay rate of the modes $\omega_{1,2}$: $\kappa_{1} = \alpha \gamma \mu_0 (H_0 + M_0/2)$ and $\kappa_{2} = \alpha \gamma \mu_0 (H_0 + M_0/2 + M_0 N)$. Since according to Eq.~(\ref{N_final}) the demagnetization factor $N$ depends on $\omega$ and $k$,  Eq.~(\ref{modes_symm_2}) represents an implicit equation for $\omega_2(k)$. The dispersion relations expressed by  Eq.~(\ref{modes_symm_1})-(\ref{modes_symm_2}) are plotted in the upper row of Fig.~\ref{fig:dispersion}. Panels (a)-(c) correspond to different angles $\theta$ between $\bm k$ and $\bm M_0$. In the absence of the ferromagnet, the superconducting S/I/S resonator hosts a highly confined electromagnetic mode found by Swihart \cite{Swihart1961}. In the presence of F layers this mode is localized within the layer
of the thickness $2(d_F+d_I+\lambda)$. The dispersion law of this mode takes the form
\begin{align}
    \Omega_s (k) = k\sqrt {\frac{d_F+d_I}{\mu_0 \varepsilon_I(d_F+d_I+\lambda)}}.
    \label{eq:swihart}
\end{align}
If there were no interaction between the Swihart mode and the magnetic excitations of the $F_1$ and $F_2$ layers, we would additionally observe two magnon modes corresponding to the Kittel frequency $\omega_K$. In fact, even in the absence of the S layers, there is a coupling between the Kittel modes in the $F_1$ and $F_2$ layers via the magnon stray fields, resulting in the splitting of their frequencies, which gives rise to the acoustic and optical eigenmodes of the coupled F/I/F system. However, the strength of the stray field is proportional to the factor $kd_F$, which is small in the considered case. Therefore, the corresponding splitting can be disregarded and cannot be resolved on the scale of Fig.~\ref{fig:dispersion}.  For this reason, to the considered accuracy, the magnon modes of the F/I/F system with $\bm M_{s1} = \bm M_{s2}$ can be viewed as degenerate with the eigenfrequency $\omega_K$. 

\begin{figure}
    \centering
    \includegraphics[width=75mm]{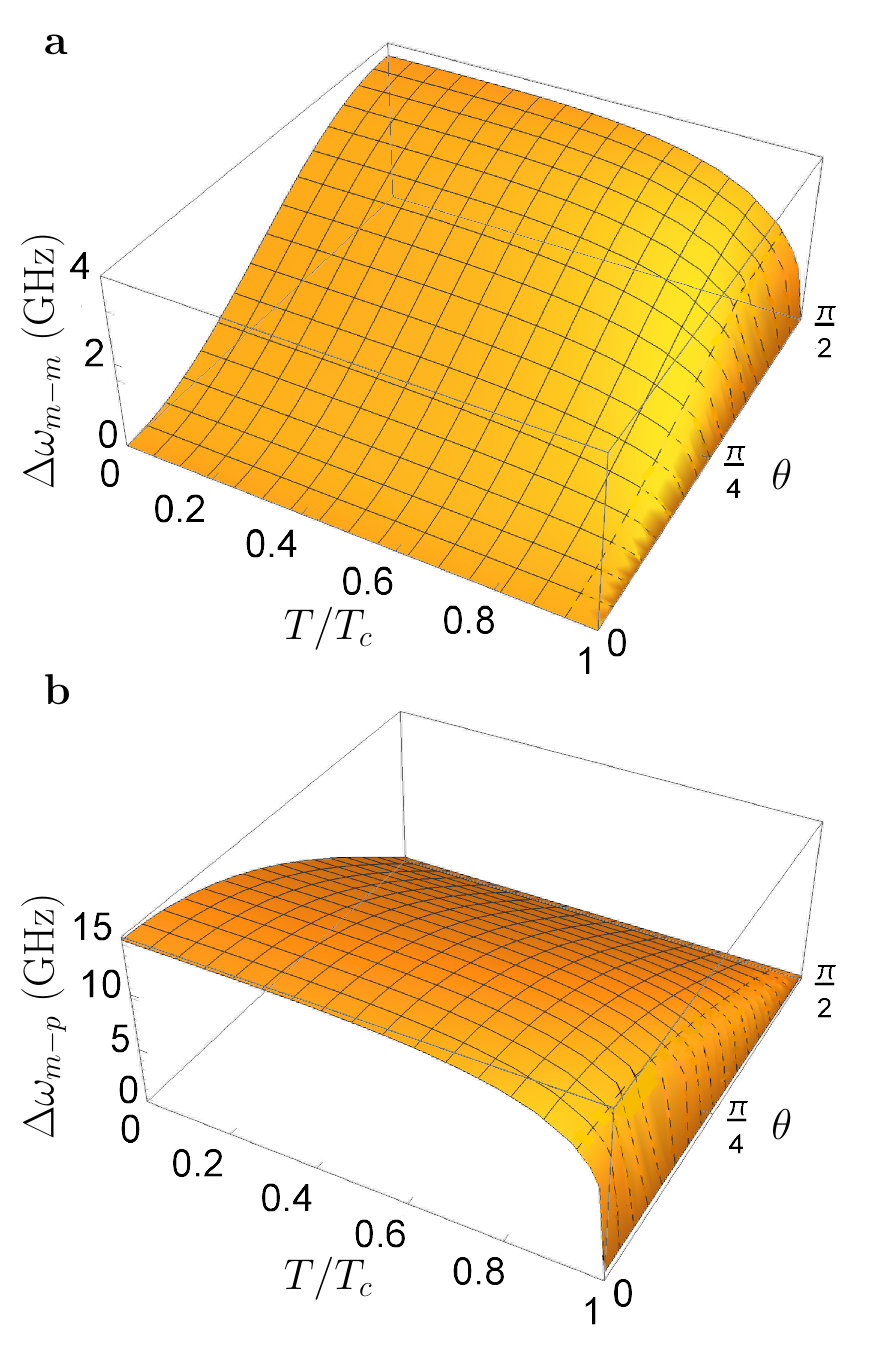}
    \caption{(a) Magnon-magnon splitting $\Delta \omega_{m-m} (k>>k_K)$ vs temperature  and $\theta$. (b) Magnon-photon splitting $\Delta \omega_{m-p} = \omega_u(k_K)-\omega_l(k_K)$ vs temperature  and $\theta$. Dashed parts of the lines indicate temperature region, where the considered model works not very well due to the contribution of the quasiparticle currents in the superconductors, see text. The parameters are the same as in Fig.~\ref{fig:dispersion} and $M_1=M_2$.}
    \label{fig:splitting}
\end{figure}

\begin{figure}
    \centering
    \includegraphics[width=88mm]{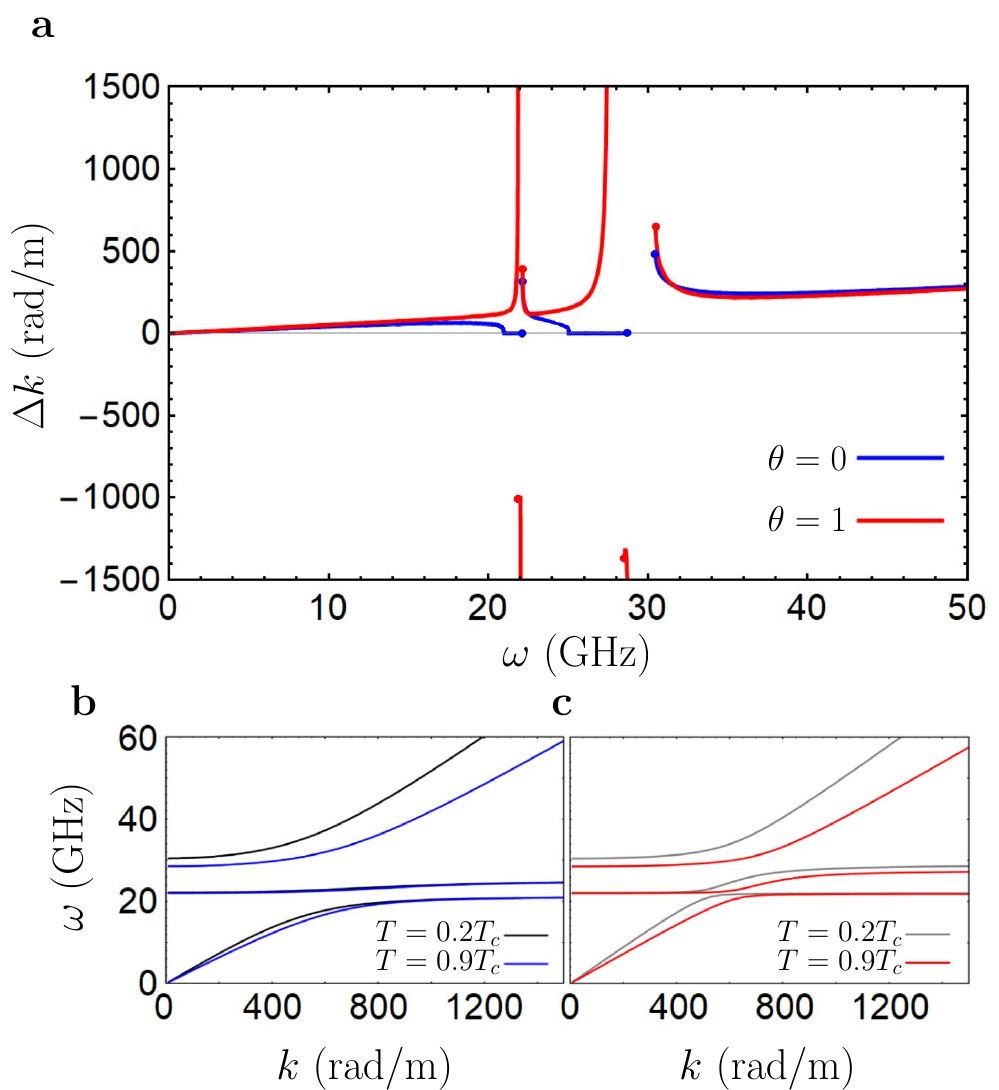}
    \caption{(a) Wave number variation $\Delta k(\omega) = k(\omega,T=0.9T_c) - k(\omega,T=0.2T_c)$ versus frequency for two different mode propagation directions $\theta$. (b)-(c) Dispersions of the eigenmodes at $\theta=0$ [(b)] and $\theta=1$ [(c)]. Black and color lines demonstrate the difference between the dispersions for both considered temperatures $T=0.2T_c$ and $T=0.9T_c$. $M_2=1.5 M_1$. The ending points of the curves between which $\Delta k$ is  undetermined are marked by the bold dots.}
    \label{fig:wavenumber_shift}
\end{figure}

However, as it is seen from the upper row of Fig.~\ref{fig:dispersion}, there is a strong coupling between the Swihart mode and one of the magnon modes, namely the acoustic mode, resulting in their anticrossing and opening a gap in the spectrum, which consists of the upper $\omega_u$ and lower $\omega_l$ excitation branches. The second magnon mode, which is the optical mode, remains uncoupled with the Swihart mode; see below. The upper $\omega_u$ and lower $\omega_l$ excitation branches, which are expressed by  Eq.~(\ref{modes_symm_2}), originate from the coupling between the Swihart mode and the acoustic magnon mode and represent magnon-photon polaritons. Equation~(\ref{modes_symm_2}) with substituting  $N$ from Eq.~(\ref{N_final}) at $d_I=0$ gives the same results for $\omega_u$ and $\omega_l$,
which were previously obtained in Ref.~\onlinecite{Silaev2023} at $\theta=0$ and in Ref.~\onlinecite{Qiu2024} at an arbitrary $\theta$ for the S/F/S heterostructure with a single ferromagnetic layer. The physical reason is that the magnetizations $\bm M_{full,1}$ and $\bm M_{full,2}$ oscillate in phase in the acoustic mode.

The magnon-photon coupling leads to the gap in the magnon-photon polariton spectrum. It can be found as $\Delta \omega_{g} = \omega_u(k=0)-\omega_l(k \gg k_K)$, where $k_K$ is the wave vector of the photon at the magnon frequency and is determined from the equation $\omega_u(k=0) = \Omega_s(k_K)$. In the limiting cases $k=0$ and $k \gg k_K$ we obtain  
\begin{align}
&\omega_u(k=0)=\gamma\mu_0\sqrt{\left(H_0+M_0 \frac{d_F}{D}\right)(H_0+M_0)}, \label{eq:omega_zero_k} \\ 
&\omega_l(k \gg k_K)=\gamma\mu_0\sqrt{\left(H_0+M_0 \frac{d_F \sin^2\theta}{D} \right)(H_0+M_0)},
\label{eq:omega_inf_k}
\end{align}
where $D=d_F+d_I+\lambda$. The coupling strength and, consequently, the splitting gap $\Delta \omega_g$ is anisotropic with respect to the mode propagation direction $\theta$. It is maximal at $\theta=0$ and vanishes at $\theta=\pi/2$. This result is in complete analogy with the anisotropy obtained in Ref.~\onlinecite{Qiu2024} for the S/F/S structures with a single F layer. For numerical estimates, we assume that the F layers are yttrium iron garnet (YIG) films with thickness $d_F=10$ nm, the thickness of the insulator layer $d_I = 10$~nm, the superconductors are NbN with London's penetration depth $\lambda(T=0)\approx 80$ nm. In addition, $\mu_0M_0=0.24$~T, $\mu_0H_0=50$~mT, $\varepsilon_{FI}=\varepsilon_I = 8\varepsilon_0$, and $\gamma=1.76\cdot10^{11}$~Hz$\cdot\mathrm{T}^{-1}$. Then the splitting gap  $\Delta \omega_g (\theta=0)$ is $\sim10$ GHz, which corresponds to the ultrastrong magnon-photon coupling, similarly to the S/F/S case \cite{Silaev2023,Qiu2024}. Different curves in each panel of Fig.~\ref{fig:dispersion} represent different temperatures. It is seen that the sensitivity of the gap to the temperature below the superconducting critical temperature is of the order of several GHz. At the same time, when we increase the temperature higher than the superconducting critical temperature $T_c$, the magnon-photon coupling becomes negligible and the spectrum is completely reconstructed. 

The spectra presented in Figs.~\ref{fig:dispersion}(a)-(c) contain another magnon mode $\omega_1 \equiv \omega_K$, which is uncoupled from the Swihart mode and remains the same as that in the absence of superconductivity. It does not depend on the temperature when the temperatures are below $T_c$. It is depicted as a red and blue line in the upper row of Fig.~\ref{fig:dispersion} to emphasize the absence of dependence on temperature. It also manifests no anisotropy with respect to the mode propagation direction $\theta$. It is the optical magnon mode of the F/I/F trilayer. Since the magnetizations of the $\rm F_1$ and $\rm F_2$ layers oscillate with the phase shift $\pi$ in the optical mode, the magnon stray fields do not generate Meissner currents in the superconductors and, therefore, this mode is not coupled to the photonic Swihart mode. 

The difference in the stray fields produced by the acoustic and optical modes is enhanced many times over by the Meissner currents. Thus the interaction between the magnon modes of the spatially separated ferromagnets is greatly enhanced as compared to the case without layers leading to 
the giant coupling between the two magnon modes. It is a new feature emerging in our S/F/I/F/S system compared to the S/F/S structures. Since the magnon-photon coupling between the magnon and the Swihart photon is essential in the vicinity of $k=k_K$, the magnon-magnon interaction can be best investigated without the hybridization of the magnon and photon modes, i.e., by analyzing the spectra at $k \to 0$ and $k \gg k_K$. At $k \gg k_K$ the demagnetization factor $N$ expressed by Eq.~(\ref{N_final}) takes the simple form
\begin{align}
N = \frac{d_F \sin^2 \theta}{2 D}.
\label{eq:N_large_k}
\end{align}
The magnon frequencies at $k \gg k_K$ are expressed by $\omega_1 = \omega_K$ and $\omega_2 = \omega_l (k \gg k_K)$. The magnon frequency splitting is 
$\Delta \omega_{m-m}(k \gg k_K) = \omega_2 - \omega_1$. It is also anisotropic, but the anisotropy is {\it opposite} to the anisotropy in the magnon-photon interaction: the maximal magnon-magnon interaction is reached at $\theta = \pi/2$, and it vanishes at $\theta=0$. It is seen in the upper row of Fig.~\ref{fig:dispersion}, where the splitting of the magnon frequencies at $k \gg k_K$ is absent in panel (a) and reaches its maximum at $\theta=\pi/2$. This maximal splitting is of the order of 10 GHz for the considered parameters, which is of the order of the bare frequency itself, and is much larger than the bare dipole-dipole interaction in the F/I/F trilayer with the same parameters. 

The dependence of $\Delta \omega_{m-m} (k>>k_K)$ on temperature and angle $\theta$ is presented in Fig.~\ref{fig:splitting}(a). It is seen that the maximal strength of the splitting is achieved at $T=0$ and $\theta =\pi/2$. At $T \to T_c$ the splitting disappears for all $\theta$. More exactly, the splitting occurs also in the normal state, but it is much smaller due to large value of the  penetration depth of microwaves $\delta \sim 1/\sqrt{2\omega \mu_0 \sigma_S} \sim 10^{-6}$ to $10^{-5}$ m  in normal state of the S layers. The temperature dependence of $\Delta \omega_{m-m} (k>>k_K)$ at $T<T_c$ results from the temperature dependence of $\lambda$. It once again indicates that the physical mechanism of magnon-magnon interaction is associated with the induction of Meissner currents in the superconductors, screening the stray magnon fields. As noted above, the approximation we are considering neglects the contribution of quasiparticle currents in the superconductors, which is justified at temperatures not very close to $T_c$. The region, where the model works not very well is shown by dashed lines in Fig.~\ref{fig:splitting}. The other possible reasons for the temperature dependence of the magnon dispersions, such as renormalization of the magnon energy and the Gilbert damping due to the excitation of thermal magnons are negligible at the cryogenic temperatures. For comparison the analogous dependence of the magnon-photon splitting gap $\Delta \omega_{m-p} = \omega_u(k_K)-\omega_l(k_K)$ on temperature and $\theta$ is presented in Fig.~\ref{fig:splitting}(b). In contrast to the magnon-magnon splitting $\Delta \omega_{m-m} (k>>k_K)$ the magnon-photon gap $\Delta \omega_{m-p}$ is maximal at $\theta=0$, that is, they manifest the opposite anisotropies with respect to the magnon propagation direction, as it was noted before.

At $k \to 0$, we also observe the giant splitting between the acoustic and the optical magnonic modes $\Delta \omega_{m-m}(k \to 0) = \omega_u(k=0)-\omega_K$. The physical reason is the same as discussed above. It is obvious that this splitting cannot manifest anisotropy with respect to the mode propagation direction $\theta$. It coincides with the value of the frequency shift of the ferromagnetic resonance frequency in a ferromagnetic insulator when sandwiched between two thick superconductors~\cite{Zhou2023}. In the considered case of two ferromagnets, this splitting is equal to the maximal value of the magnon-magnon splitting at $k \gg k_K$, which is reached at $\theta = \pi/2$. Therefore, in general, the anisotropy of the magnon-magnon coupling depends on the value of the magnon wave vector $k$. 

\subsection{Anisotropic giant magnon splitting and magnon-polaritons. Different F-layers.}

Now we generalize our consideration to the asymmetric case with $M_1\neq M_2$. Then the demagnetization tensor is still expressed by Eq.~(\ref{N_final}), but the eigenfrequencies, which are to be calculated from Eq.~(\ref{M_matrix}), take the form: 
\begin{align}
    \omega_{1,2}^2=&\frac{1}{2}\left[ \omega_{01}^2 + \omega_{02}^2  \right. \nonumber \\
    &\left. \pm\sqrt{(\omega_{01}^2 + \omega_{02}^2)^2-4 \omega_{01}^2 \omega_{02}^2(1-L^2)} \right],
    \label{modes_asymm}
\end{align}
where 
\begin{align}
    \omega_{01(2)} = \gamma \mu_0 \sqrt{(H_0 + M_{1(2)})(H_0 + N M_{1(2)})}
    \label{modes_asymm_0}
\end{align}
and 
\begin{align}
    L = N \sqrt{\frac{M_1 M_2}{(H_0 + N M_1)(H_0 + N M_2)}}.
    \label{L}
\end{align}
The spectra with $M_1 \neq M_2$ are presented in the bottom row of Fig.~\ref{fig:dispersion}. When $T>T_c$, they are just two uncoupled Kittel modes $\omega_{K,1(2)} = \gamma \mu_0 \sqrt{H_0(H_0+M_{1(2)})}$, which are shown in Figs.~\ref{fig:dispersion}(d)-(f) by the dashed lines. These results can be obtained from Eq.~(\ref{modes_asymm}) in the limit $\lambda \to \infty$ corresponding to $N \to 0$. When the temperature drops below $T_c$, the interaction between the magnon modes and between the magnon and Swihart modes via the Meissner supercurrents is switched on, and the giant shift of the uncoupled frequencies $\omega_{K,1(2)}$ appears. With $M_1 \neq M_2$, both magnon modes, acoustic and optical, are coupled to the photon because the magnon stray fields, in this case, cannot be fully canceled even for the optical mode. As a result of the coupling to the photon, both magnetic modes depend on $k$ in this case, as is seen in Figs.~\ref{fig:dispersion}(d)-(e). The magnonic modes become dispersionless only at $\theta = \pi/2$, when the coupling between the magnons and the photon vanishes, as discussed above.

With $M_1 \neq M_2$, the same key statements in the symmetric configuration regarding the anisotropy of magnon-magnon and magnon-photon interactions remain valid. The magnon-photon coupling is maximal at $\theta = 0$ and disappears at $\theta = \pi/2$. For the magnon-magnon coupling at $k \gg k_K$ the situation is opposite, it is maximal at $\theta =\pi/2$ and vanishes at $\theta = 0$. It is clearly seen in the asymptotic behavior of the magnon modes at $k \gg k_K$. Namely, both magnonic modes tend to their uncoupled values $\omega_{K,1(2)}$ at $\theta = 0$, but there is a giant frequency shift of the order of several GHz of both magnonic modes at $\theta=\pi/2$.

\subsection{Temperature tuning of magnon wavenumbers}

The strong dependence of the frequency spectra of the collective modes on the temperature allows one to implement an efficient tuning of their wave number in the S/F/I/F/S heterostructures controlled by the temperature, which can be important for applications requiring essential and easily controllable phase shifts of propagating excitations. The magnonic phase shifters are responsible for the phase modulation and the interference output, which is critical for magnonic logic gates. Various phase-shifting mechanisms have been proposed in the literature. The main approaches include static phase shifters based on the domain walls \cite{Hertel2004,Bayer2005,Duerr2012}, magnetic defects \cite{Louis2016,Baumgaertl2018}, magnonic crystals \cite{Zhu2014}, nanomagnets~\cite{Au2012}, and dynamic phase shifters, which could be controlled in real time by an external signal, what can be implemented via an applied magnetic field \cite{Kostylev2005,Ustinov2013}, electric currents \cite{Hansen2009}, spin-polarized currents \cite{Wang2014,Chen2015,Zhang2020} or electric fields \cite{Rana2018,Liu2011,Zhang2014,Wang2018_2,Nikitin2017_3,Nikitin2015}. 

Here, we suggest implementing a tuning of the wave number by variations of temperature below $T_c$. Let us consider the frequency spectra of the excitation modes of the S/F/I/F/S heterostructure at two different temperatures below $T_c$. The examples corresponding to $T=0.2T_c$ and $T=0.9T_c$ are presented in Figs.~\ref{fig:wavenumber_shift}(b) and (c). The wavenumber variation $\Delta k(\omega) = k(\omega,T=0.9T_c) - k(\omega,T=0.2T_c)$  versus frequency is plotted in Fig.~\ref{fig:wavenumber_shift}(a) for two different mode propagation directions $\theta$.  It is seen that large $\Delta k \sim 1000$ rad/m can be obtained in a wide range of frequencies of the order of tens of GHz. The effect exists for a wide range of $\theta$ except for $\theta$ close to $\pi/2$ because, in this case, the magnon-photon coupling that renders the magnon modes dependent on $k$ disappears. Moreover, at finite values of $\theta$, one can obtain a huge positive (negative) value of $\Delta k$ in some narrow regions of $\omega$ because at such frequencies, the low(high)-temperature excitation mode has finite $k$ and for the high(low)-temperature mode $k\to \infty$; see the frequency spectra at the low temperature and high temperature shown in Figs.~\ref{fig:wavenumber_shift}(b)-(c). Also, there are frequency intervals in Fig.~\ref{fig:wavenumber_shift}(a), where $\Delta k(\omega)$ is indetermined. They correspond to the gaps in the spectra presented in Figs.~\ref{fig:wavenumber_shift}(b) and (c). In Fig.~\ref{fig:wavenumber_shift}(a) the ending points of the curves between which $\Delta k(\omega)$ is undetermined are marked by bold dots.

\section{Conclusions}
\label{conclusions}

In conclusion, a theoretical approach is developed for calculating the frequency spectra of the magnetic excitations in heterostructures containing several ferromagnetic layers sandwiched between two superconductors. It is based on the solution of the coupled system of Maxwell equations and the Landau-Lifshitz-Gilbert equation. The developed method is applied to finding the spectra of magnetic excitations in S/F/I/F/S heterostructure. It is shown that in such a system, magnon excitations of the F-layers are accompanied by Meissner supercurrents in the S-layers, which strongly enhance the magnon stray fields and, thus, mediate the ultra-strong interaction between the magnon modes of the ferromagnetic layers. As a result of the magnon-magnon interaction of such composite excitations, the spectrum of the magnetic excitations, which consists of acoustic and optical modes, manifests several properties that are interesting from the fundamental point of view and promising from the point of view of applications. 

(i) In the general case, both modes interact with the electromagnetic Swihart mode of the superconducting resonator. The interaction of the acoustic mode with the Swihart mode is ultra-strong, in full agreement with the results obtained earlier for a system with a single ferromagnetic layer \cite{Silaev2023,Qiu2024}. The interaction of the optical mode with the Swihart mode is much weaker due to the weakening of the stray magnetic fields in this mode and completely disappears in the case of identical F-layers with $M_1=M_2$.

(ii) The interaction of both magnon modes with the Swihart mode is anisotropic. Its strength is maximum at zero angle $\theta$ between the direction of the equilibrium magnetization of the magnets and the excitation wave vector $\bm k$ and disappears at $\theta=\pi/2$.

(iii) The magnon-magnon interaction of composite magnetic excitations without an admixture of interaction with the Swihart mode is realized if $k$ is far from the region of interaction of magnons with the Swihart mode $k \sim k_K$. It is also ultra-strong. The rearrangement of the magnon frequencies resulting from this interaction is of the order of the bare frequency itself. At large $k \gg k_K$, the magnon-magnon interaction is also anisotropic, but its anisotropy is opposite to the anisotropy of the magnon-photon interaction with the Swihart mode: the interaction strength is maximal at $\theta=\pi/2$ and vanishes at $\theta=0$. 

(iv) Both the magnon-magnon interaction strength and the interaction strength of magnon excitations with the Swihart mode depend very strongly on the temperature, which is physically a consequence of the temperature dependence of the penetration depth of the magnetic field in the superconductors. It allows to implement an effective tuning of the wave number in the S/F/I/F/S heterostructures controlled by temperature in a wide range
of frequencies of the order of tens of GHz, which can be important for applications requiring essential and easily controllable phase shifts of propagating excitations.

\begin{acknowledgments}
The authors are grateful to Dr. M. A. Silaev for suggesting the initial idea of this work. V.M.G., G.A.B. and I.V.B. acknowledge the support from Theoretical Physics and Mathematics Advancement Foundation “BASIS” via the project No. 23-1-1-51-1. The development of the approach and calculation of magnonic spectra was supported by the Russian Science Foundation via the RSF project No. 22-42-04408 and studying of temperature tuning of magnon wave number was supported by Grant from the Ministry of Science and Higher Education of the Russian Federation No. 075-15-2024-632. T.Y. is financially supported by the National Key Research and Development Program of China under Grant No. 2023YFA1406600 and the National Natural Science Foundation of China under Grant No. 12374109.
\end{acknowledgments}

\section*{Appendix: derivation of the demagnetization tensor}
Here we present the detailed derivation of the expression for the demagnetization tensor $\hat N$ [Eq. (\ref{N_final})]. Let us explicitly write the boundary conditions for the Maxwell equations. The continuity of the components $E_y$ and $E_z$, obtained from Eq. (\ref{Maxwell_solutions}), at the four interfaces $x=\pm d_I$, $x=\pm (d_I+d_F)$ leads to the following two systems, respectively:
\begin{align}
    &S_{1y}e^{iBd_F}=F_{1y}^+e^{iCd_F}+F_{1y}^-e^{-iCd_F}+\dfrac{\omega\mu_0}{C^2}k_z\tilde M_{1x},\nonumber\\
    &I_{y}^+e^{iAd_I}+ I_{y}^-e^{-iAd_I}=F_{1y}^++F_{1y}^-+\dfrac{\omega\mu_0}{C^2}k_z\tilde M_{1x},\nonumber\\
    &I_{y}^+e^{-iAd_I}+ I_{y}^-e^{iAd_I}=F_{2y}^++F_{2y}^-+\dfrac{\omega\mu_0}{C^2}k_z\tilde M_{2x},\nonumber\\
    &S_{2y}e^{iBd_F}=F_{2y}^+e^{-iCd_F}+F_{2y}^-e^{iCd_F}+\dfrac{\omega\mu_0}{C^2}k_z\tilde M_{2x},
    \label{boundary_Ey}
\end{align}
and
\begin{align}
    &S_{1z}e^{iBd_F}=F_{1z}^+e^{iCd_F}+F_{1z}^-e^{-iCd_F}-\dfrac{\omega\mu_0}{C^2}k_y\tilde M_{1x},\nonumber\\
    &I_{z}^+e^{iAd_I}+ I_{z}^-e^{-iAd_I}=F_{1z}^++F_{1z}^--\dfrac{\omega\mu_0}{C^2}k_y\tilde M_{1x},\nonumber\\
    &I_{z}^+e^{-iAd_I}+ I_{z}^-e^{iAd_I}=F_{2z}^++F_{2z}^--\dfrac{\omega\mu_0}{C^2}k_y\tilde M_{2x},\nonumber\\
    &S_{2z}e^{iBd_F}=F_{2z}^+e^{-iCd_F}+F_{2z}^-e^{iCd_F}-\dfrac{\omega\mu_0}{C^2}k_y\tilde M_{2x}.
    \label{boundary_Ez}
\end{align}
The continuity of $H_z$, which is obtained from  Eq. (\ref{H_final}) with the use of Eq. (\ref{Maxwell_solutions}), gives us
\begin{widetext}

\begin{align}
    &iBS_{1y}e^{iBd_F}=\dfrac{1}{C^2}\left(iC\alpha_F(F_{1y}^+e^{iCd_F}-F_{1y}^-e^{-iCd_F})+iC\dfrac{K_1}{2}(F_{1z}^+e^{iCd_F}-F_{1z}^-e^{-iCd_F})+i\omega\mu_0\dfrac{K_1}{2}\tilde M_{1y}\right),\nonumber\\
    &\dfrac{i}{A}\left(\alpha_I(I_{y}^+e^{iAd_I}- I_{y}^-e^{-iAd_I})+\dfrac{K_1}{2}(I_{z}^+e^{iAd_I}- I_{z}^-e^{-iAd_I})\right)\nonumber\\
    &=\dfrac{1}{C^2}\left(iC\alpha_F(F_{1y}^+-F_{1y}^-)+iC\dfrac{K_1}{2}(F_{1z}^+-F_{1z}^-)+i\omega\mu_0\dfrac{K_1}{2}\tilde M_{1y}\right),\nonumber\\
    &\dfrac{i}{A}\left(\alpha_I(I_{y}^+e^{-iAd_I}- I_{y}^-e^{iAd_I})+\dfrac{K_1}{2}(I_{z}^+e^{-iAd_I}- I_{z}^-e^{iAd_I})\right)\nonumber\\
    &=\dfrac{1}{C^2}\left(iC\alpha_F(F_{2y}^+-F_{2y}^-)+iC\dfrac{K_1}{2}(F_{2z}^+-F_{2z}^-)+i\omega\mu_0\dfrac{K_1}{2}\tilde M_{2y}\right),\nonumber\\
    &-iBS_{2y}e^{iBd_F}=\dfrac{1}{C^2}\left(iC\alpha_F(F_{2y}^+e^{-iCd_F}-F_{2y}^-e^{iCd_F})+iC\dfrac{K_1}{2}(F_{2z}^+e^{-iCd_F}-F_{2z}^-e^{iCd_F})+i\omega\mu_0\dfrac{K_1}{2}\tilde M_{2y}\right),
     \label{boundary_Hz}
\end{align}
and the continuity of $H_y$ analogously leads to
\begin{align}
    &iBS_{1z}e^{iBd_F}=\dfrac{1}{C^2}\left(iC\beta_F(F_{1z}^+e^{iCd_F}-F_{1z}^-e^{-iCd_F})+iC\dfrac{K_1}{2}(F_{1y}^+e^{iCd_F}-F_{1y}^-e^{-iCd_F})+i\omega\mu_0\beta_F\tilde M_{1y}\right),\nonumber\\
    &\dfrac{i}{A}\left(\beta_I(I_{z}^+e^{iAd_I}- I_{z}^-e^{-iAd_I})+\dfrac{K_1}{2}(I_{y}^+e^{iAd_I}- I_{y}^-e^{-iAd_I})\right)\nonumber\\
    &=\dfrac{1}{C^2}\left(iC\beta_F(F_{1z}^+-F_{1z}^-)+iC\dfrac{K_1}{2}(F_{1y}^+-F_{1y}^-)+i\omega\mu_0\beta_F\tilde M_{1y}\right),\nonumber\\
    &\dfrac{i}{A}\left(\beta_I(I_{z}^+e^{-iAd_I}- I_{z}^-e^{iAd_I})+\dfrac{K_1}{2}(I_{y}^+e^{-iAd_I}- I_{y}^-e^{iAd_I})\right)\nonumber\\
    &=\dfrac{1}{C^2}\left(iC\beta_F(F_{2z}^+-F_{2z}^-)+iC\dfrac{K_1}{2}(F_{2y}^+-F_{2y}^-)+i\omega\mu_0\beta_F\tilde M_{2y}\right),\nonumber\\
   & -iBS_{2z}e^{iBd_F}=\dfrac{1}{C^2}\left(iC\beta_F(F_{2z}^+e^{-iCd_F}-F_{2z}^-e^{iCd_F})+iC\dfrac{K_1}{2}(F_{2y}^+e^{-iCd_F}-F_{2y}^-e^{iCd_F})+i\omega\mu_0\beta_F\tilde M_{2y}\right).
     \label{boundary_Hy}
\end{align}

Let us introduce symmetrized and antisymmetrized amplitudes of the electric field in the I and F layers:
 \begin{align}
 & I_{y,z}^{s}=I_{y,z}^++ I_{y,z}^-, \nonumber\\
 & I_{y,z}^{a}=I_{y,z}^+- I_{y,z}^-, \nonumber\\
& F_{1(2)y,z}^{s}=F_{1(2)y,z}^++ F_{1(2)y,z}^-, \nonumber\\
& F_{1(2)y,z}^{a}=F_{1(2)y,z}^+- F_{1(2)y,z}^-.
\label{symm_ampl}
 \end{align}
From now on, we shall expand all the boundary conditions up to the first order with respect to $|Ad_I|$ and $ |Cd_{F}|$, which was motivated in the main text. The second and the third equations from the systems (\ref{boundary_Ey}) and (\ref{boundary_Ez}), rewritten in terms of the amplitudes (\ref{symm_ampl}), in the linear order with respect to $|Ad_I|$ give us:
\begin{align}
& I_{y}^{s}=\dfrac{1}{2}(F_{1y}^{s}+F_{2y}^{s}+\dfrac{\omega\mu_0}{C^2}k_z(\tilde M_{1x}+\tilde M_{2x})), \nonumber\\
& I_{y}^{a}=\dfrac{1}{2iAd_I}(F_{1y}^{s}-F_{2y}^{s}+\dfrac{\omega\mu_0}{C^2}k_z(\tilde M_{1x}-\tilde M_{2x})),\nonumber\\
& I_{z}^{s}=\dfrac{1}{2}(F_{1z}^{s}+F_{2z}^{s}-\dfrac{\omega\mu_0}{C^2}k_y(\tilde M_{1x}+\tilde M_{2x})), \nonumber\\
& I_{z}^{a}=\dfrac{1}{2iAd_I}(F_{1z}^{s}-F_{2z}^{s}-\dfrac{\omega\mu_0}{C^2}k_y(\tilde M_{1x}-\tilde M_{2x})).
\label{I_final}
 \end{align}
 Then we take the rest two equations from the system (\ref{boundary_Ey}) and the whole system (\ref{boundary_Hz}) and get rid of the amplitudes $S_{1(2)y}$. In terms of the new amplitudes (\ref{symm_ampl}) and in the first order with respect to $|Ad_I|$ and $ |Cd_{F}|$ we get:
 \begin{align}
    & B(F_{1y}^{s}+iCd_FF_{1y}^{a}+\dfrac{\omega\mu_0}{C^2}k_z\tilde M_{1x})=\dfrac{1}{C}\left(\alpha_F(F_{1y}^{a}+iCd_FF_{1y}^{s})+\dfrac{K_1}{2}(F_{1z}^{a}+iCd_FF_{1z}^{s})\right)+\dfrac{\omega\mu_0K_1}{2C^2}\tilde M_{1y},\nonumber\\
     &\dfrac{1}{A}\left(\alpha_I \dfrac{F_{1y}^{s}-F_{2y}^{s}+\dfrac{\omega\mu_0}{C^2}k_z(\tilde M_{1x}-\tilde M_{2x})}{2iAd_I}+\dfrac{K_1}{2}\dfrac{F_{1z}^{s}-F_{2z}^{s}-\dfrac{\omega\mu_0}{C^2}k_y(\tilde M_{1x}-\tilde M_{2x})}{2iAd_I}\right)\nonumber\\
     &=\dfrac{1}{C}\left( \alpha_FF_{1y}^{a}+\dfrac{K_1}{2}F_{1z}^{a}\right)+\dfrac{\omega\mu_0K_1}{2C^2}\tilde M_{1y},\nonumber\\
     &\dfrac{1}{A}\left(\alpha_I \dfrac{F_{1y}^{s}-F_{2y}^{s}+\dfrac{\omega\mu_0}{C^2}k_z(\tilde M_{1x}-\tilde M_{2x})}{2iAd_I}+\dfrac{K_1}{2}\dfrac{F_{1z}^{s}-F_{2z}^{s}-\dfrac{\omega\mu_0}{C^2}k_y(\tilde M_{1x}-\tilde M_{2x})}{2iAd_I}\right)\nonumber\\
     &=\dfrac{1}{C}\left( \alpha_FF_{2y}^{a}+\dfrac{K_1}{2}F_{2z}^{a}\right)+\dfrac{\omega\mu_0K_1}{2C^2}\tilde M_{2y},\nonumber\\
    &-B(F_{2y}^{s}-iCd_FF_{2y}^{a}+\dfrac{\omega\mu_0}{C^2}k_z\tilde M_{2x})=\dfrac{1}{C}\left(\alpha_F(F_{2y}^{a}-iCd_FF_{2y}^{s})+\dfrac{K_1}{2}(F_{2z}^{a}-iCd_FF_{2z}^{s})\right)+\dfrac{\omega\mu_0K_1}{2C^2}\tilde M_{2y}.
    \label{boundary1_final}
 \end{align}
 Repeating the similar procedure for the systems (\ref{boundary_Ez}) and (\ref{boundary_Hy}), we obtain:
 \begin{align}
   &  B(F_{1z}^{s}+iCd_FF_{1z}^{a}-\dfrac{\omega\mu_0}{C^2}k_y\tilde M_{1x})=\dfrac{1}{C}\left(\beta_F(F_{1z}^{a}+iCd_FF_{1z}^{s})+\dfrac{K_1}{2}(F_{1y}^{a}+iCd_FF_{1y}^{s})\right)+\dfrac{\omega\mu_0\beta_F}{C^2}\tilde M_{1y},\nonumber\\
  &   \dfrac{1}{A}\left(\beta_I \dfrac{F_{1z}^{s}-F_{2z}^{s}-\dfrac{\omega\mu_0}{C^2}k_y(\tilde M_{1x}-\tilde M_{2x})}{2iAd_I}+\dfrac{K_1}{2}\dfrac{F_{1y}^{s}-F_{2y}^{s}+\dfrac{\omega\mu_0}{C^2}k_z(\tilde M_{1x}-\tilde M_{2x})}{2iAd_I}\right)=\dfrac{1}{C}\left( \beta_FF_{1z}^{a}+\dfrac{K_1}{2}F_{1y}^{a}\right)\nonumber\\
  &+\dfrac{\omega\mu_0\beta_F}{C^2}\tilde M_{1y},\nonumber\\
 &    \dfrac{1}{A}\left(\beta_I \dfrac{F_{1z}^{s}-F_{2z}^{s}-\dfrac{\omega\mu_0}{C^2}k_y(\tilde M_{1x}-\tilde M_{2x})}{2iAd_I}+\dfrac{K_1}{2}\dfrac{F_{1y}^{s}-F_{2y}^{s}+\dfrac{\omega\mu_0}{C^2}k_z(\tilde M_{1x}-\tilde M_{2x})}{2iAd_I}\right)=\dfrac{1}{C}\left( \beta_FF_{2z}^{a}+\dfrac{K_1}{2}F_{2y}^{a}\right)\nonumber\\
 &+\dfrac{\omega\mu_0\beta_F}{C^2}\tilde M_{2y},\nonumber\\
&    -B(F_{2z}^{s}-iCd_FF_{2z}^{a}-\dfrac{\omega\mu_0}{C^2}k_y\tilde M_{2x})=\dfrac{1}{C}\left(\beta_F(F_{2z}^{a}-iCd_FF_{2z}^{s})+\dfrac{K_1}{2}(F_{2y}^{a}-iCd_FF_{2y}^{s})\right)+\dfrac{\omega\mu_0\beta_F}{C^2}\tilde M_{2y}.    \label{boundary2_final}
 \end{align}
 
 The systems (\ref{boundary1_final}) and (\ref{boundary2_final}) can be written in the form of the following matrix equation on $\hat F=(F_{1y}^s, F_{1y}^a, F_{2y}^s, F_{2y}^a,F_{1z}^s, F_{1z}^a, F_{2z}^s, F_{2z}^a)^T$:
 \begin{align}
     \hat K\hat F=\hat M,
     \label{boundary_cond}
 \end{align}
 where
 \begin{align}
     \hat K=\left(\begin{array}{cccccccc}
B-id_F\alpha_F & iBCd_{F}-\dfrac{\alpha_F}{C} & 0 & 0 &  -\dfrac{id_FK_1}{2} & -\dfrac{K_1}{2C} & 0 & 0\\ \dfrac{\alpha_I}{A}
 &-\dfrac{iAd_I\alpha_F}{C}&-\dfrac{\alpha_I}{A}&-\dfrac{iAd_I\alpha_F}{C}&\dfrac{K_1}{2A}&-\dfrac{iAd_IK_1}{2C}&-\dfrac{K_1}{2A}&-\dfrac{iAd_IK_1}{2C}\\0&-\dfrac{\alpha_F}{C}&0&\dfrac{\alpha_F}{C}&0&-\dfrac{K_1}{2C}&0&\dfrac{K_1}{2C}\\0&0&-B+id_F\alpha_F&iBCd_{F}-\dfrac{\alpha_F}{C}&0&0&\dfrac{id_FK_1}{2}&-\dfrac{K_1}{2C}\\-\dfrac{id_FK_1}{2}&-\dfrac{K_1}{2C}&0&0&B-id_F\beta_F&iBCd_{F}-\dfrac{\beta_F}{C}&0&0\\\dfrac{K_1}{2A}&-\dfrac{iAd_IK_1}{2C}&-\dfrac{K_1}{2A}&-\dfrac{iAd_IK_1}{2C}&\dfrac{\beta_I}{A}&-\dfrac{iAd_I\beta_F}{C}&-\dfrac{\beta_I}{A}&-\dfrac{iAd_I\beta_F}{C}\\0&-\dfrac{K_1}{2C}&0&\dfrac{K_1}{2C}&0&-\dfrac{\beta_F}{C}&0&\dfrac{\beta_F}{C}\\0&0&\dfrac{id_FK_1}{2}&-\dfrac{K_1}{2C}&0&0&-B+id_F\beta_F&iBCd_{F}-\dfrac{\beta_F}{C}
\end{array}\right),
\end{align}
 \begin{align}
     \hat M=\dfrac{k^2\omega\mu_0}{C^2}\left(\begin{array}{c}
    -\dfrac{B}{k}\cos\theta \tilde M_{1x}+\dfrac{\sin2\theta}{2}\tilde M_{1y}\\ A\left( -\dfrac{\cos \theta}{k}
 (\tilde M_{1x}-\tilde M_{2x})+id_I\dfrac{\sin 2\theta}{2}(\tilde M_{1y}+\tilde M_{2y})\right)\\\dfrac{\sin2\theta}{2}(\tilde M_{1y}-\tilde M_{2y})\\\dfrac{B}{k}\cos\theta\tilde M_{2x}+\dfrac{\sin2\theta}{2}\tilde M_{2y}\\\dfrac{B}{k}\sin\theta\tilde M_{1x}+\dfrac{\beta_F}{k^2}\tilde M_{1y}\\A\left(\dfrac{\sin \theta}{k}(\tilde M_{1x}-\tilde M_{2x})+i\dfrac{d_I\beta_F}{k^2}(\tilde M_{1y}+\tilde M_{2y})\right)\\\dfrac{\beta_F}{k^2}(\tilde M_{1y}-\tilde M_{2y})\\-\dfrac{B}{k}\sin\theta\tilde M_{2x}+\dfrac{\beta_F}{k^2}\tilde M_{2y}
 \end{array}\right),
 \end{align}
(in each of the systems (\ref{boundary1_final}), (\ref{boundary2_final}) we have replaced the second and the third lines with their sum divided by 2 and their difference divided by $2iAd_I$, respectively).

Now our goal is to derive the relation between $\hat H$ and $\hat {\tilde M}$. First, we can express the components of $\hat H$ via the components of $\hat F$ and $\hat {\tilde M}$ using Eq.~(\ref{H_final}) and Eq.~(\ref{Maxwell_solutions}):
\begin{align}
   & \tilde H_{x1(2)}=\dfrac{k}{\omega\mu_0}(F_{1(2)z}^s\sin\theta-F_{1(2)y}^s\cos\theta)-\dfrac{k_F^2}{C^2}\tilde M_{1(2)x},\nonumber\\
   & \tilde H_{y1(2)}=-\dfrac{1}{C\omega\mu_0}\left(F_{1(2)z}^a\beta_F+F_{1(2)y}^a\dfrac{K_1}{2}\right)-\dfrac{\beta_F}{C^2}\tilde M_{1(2)y},
    \label{H_FM}
\end{align}
where we have written the expressions for $E_{y,z}$ and $\partial_x E_{z}$ at the F1(2)/I boundaries, as the magnetic field can be considered independent of $x$ in each of the F layers. Eq. (\ref{H_FM}) can be rewritten in the matrix form:
\begin{align}
    \hat H=\hat \Gamma\hat F+\hat\Gamma_M\hat {\tilde M},
    \label{matrix H_FM}
\end{align}
where
\begin{align}
    &\hat\Gamma=\dfrac{1}{\omega\mu_0}\left(\begin{array}{cccccccc}
        -k\cos\theta & 0&0&0&k\sin\theta & 0&0&0 \\
        0&0 &-k\cos\theta & 0&0&0 &k\sin\theta & 0\\
        0&-\dfrac{K_1}{2C}& 0&0&0&-\dfrac{\beta_F}{C}&0&0\\
        0&0&0&-\dfrac{K_1}{2C}& 0&0&0&-\dfrac{\beta_F}{C}
    \end{array}\right),\nonumber\\
    &\hat\Gamma_M=\left(\begin{array}{cccc}
        -\dfrac{k_F^2}{C^2} & 0&0&0 \\
         0& -\dfrac{k_F^2}{C^2} & 0&0\\
         0&0&-\dfrac{\beta_F}{C^2}&0\\
          0&0&0&-\dfrac{\beta_F}{C^2}
    \end{array}\right).
\end{align}
Finally, in Eq. (\ref{matrix H_FM}) we can substitute $\hat F$ expressed via $\hat {\tilde M}$:
\begin{align}
    \hat F=\hat K^{-1}\hat M=\hat K^{-1}\hat M_t\hat {\tilde M}, ~~\hat M_t=\dfrac{\omega\mu_0}{C^2}\left(\begin{array}{cccc}
        -Bk\cos\theta &0&\dfrac{K_1}{2}&0  \\
         -Ak\cos\theta& Ak\cos\theta&iAd_I\dfrac{K_1}{2}&iAd_I\dfrac{K_1}{2}\\
         0&0&\dfrac{K_1}{2}&-\dfrac{K_1}{2}\\
         0&Bk\cos\theta &0&\dfrac{K_1}{2}\\
         Bk\sin\theta&0&\beta_F&0\\
         Ak\sin\theta&-Ak\sin\theta&iAd_I\beta_F&iAd_I\beta_F\\
         0&0&\beta_F&-\beta_F\\
         0&-Bk\sin\theta&0&\beta_F
    \end{array} \right).
\end{align}
Then, comparing Eq. (\ref{matrix H_FM}) with Eq. (\ref{H_M}), we get
\begin{align}
    \hat N=-(\hat\Gamma\hat K^{-1}\hat M_t+\hat\Gamma_M),
\end{align}
which leads to the final expression for the demagnetization tensor:
\begin{align}
     \hat N=\left(\begin{array}{cccc}
        1  & 0&0&0 \\
         0 &1&0&0\\
         0&0&N&N\\
         0&0&N&N
     \end{array}\right),~~~
     N=\dfrac{Bd_{F}(k_I^2 k_F^2\alpha-Bk^2\chi\sin^2\theta)}{2\alpha(Bk^2\chi-k_I^2 k_F^2\alpha)},
 \end{align}
 $\alpha=i+B(d_F+d_I)$, $\chi=d_Fk_I^2+d_Ik_F^2$. Here all the elements of the tensor $\hat N$ are written up to the leading order with respect to $|Ad_I|$ and $ |Cd_{F}|$.
  \end{widetext}

\bibliography{SFIFS}

\end{document}